\documentclass[12pt]{article}
\usepackage{amssymb}

\newcommand{\D}{\Delta}

\def\be{\begin{equation}}
\def\ee{\end{equation}}
\def\bea{\begin{eqnarray}}
\def\eea{\end{eqnarray}}
\def\ba{\begin{array}}
\def\ea{\end{array}}
\def\bi{\begin{itemize}}
\def\ei{\end{itemize}}

\def\Journal#1#2#3#4{{#1} {\bf #2}, #3 (#4)}

\def\NPB{{\em Nucl. Phys.} B}

\def\PRD{{\em Phys. Rev.} D}

\catcode`\@=11
\def\@citex[#1]#2{%
\if@filesw \immediate \write \@auxout {\string \citation {#2}}\fi
\@tempcntb\m@ne \let\@h@ld\relax \def\@citea{}%
\@cite{%
  \@for \@citeb:=#2\do {%
    \@ifundefined {b@\@citeb}%
      {\@h@ld\@citea\@tempcntb\m@ne{\bf ?}%
      \@warning {Citation `\@citeb ' on page \thepage \space undefined}}%
      {\@tempcnta\@tempcntb \advance\@tempcnta\@ne%
      \@tempcntb\number\csname b@\@citeb \endcsname \relax%
      \ifnum\@tempcnta=\@tempcntb 
        \ifx\@h@ld\relax%
          \edef \@h@ld{\@citea\csname b@\@citeb\endcsname}%
        \else%
          \edef\@h@ld{\ifmmode{-}\else--\fi\csname b@\@citeb\endcsname}%
        \fi%
      \else
        \@h@ld\@citea\csname b@\@citeb \endcsname%
        \let\@h@ld\relax%
      \fi}%
    \def\@citea{,\penalty\@highpenalty\,}%
  }\@h@ld
}{#1}}

\def\@citeb#1#2{{[#1]\if@tempswa , #2\fi}}
\def\@citeu#1#2{{$^{#1}$\if@tempswa , #2\fi }}
\def\@citep#1#2{{#1\if@tempswa , #2\fi}}

\def\bcites{         
        \catcode`\@=11
        \let\@cite=\@citeb
        \catcode`\@=12
}

\def\upcites{         
        \catcode`\@=11
        \let\@cite=\@citeu
        \catcode`\@=12
}

\def\plaincites{      
        \catcode`\@=11
        \let\@cite=\@citep
        \catcode`\@=12
}

\newcount\hour
\newcount\minute
\newtoks\amorpm
\hour=\time\divide\hour by 60
\minute=\time{\multiply\hour by 60 \global\advance\minute by-\hour}
\edef\standardtime{{\ifnum\hour<12 \global\amorpm={am}%
        \else\global\amorpm={pm}\advance\hour by-12 \fi
        \ifnum\hour=0 \hour=12 \fi
        \number\hour:\ifnum\minute<10 0\fi\number\minute\the\amorpm}}
\edef\militarytime{\number\hour:\ifnum\minute<10 0\fi\number\minute}

\def\draftlabel#1{{\@bsphack\if@filesw {\let\thepage\relax
   \xdef\@gtempa{\write\@auxout{\string
      \newlabel{#1}{{\@currentlabel}{\thepage}}}}}\@gtempa
   \if@nobreak \ifvmode\nobreak\fi\fi\fi\@esphack}
        \gdef\@eqnlabel{#1}}
\def\@eqnlabel{}
\def\@vacuum{}
\def\marginnote#1{}
\def\draftmarginnote#1{\marginpar{\raggedright\scriptsize\tt#1}}
\def\draft{
        \pagestyle{plain}
        \overfullrule=2pt
        \oddsidemargin -.5truein
        \def\@oddhead{\sl \phantom{\today\quad\militarytime} \hfil
        \smash{\Large\sl DRAFT} \hfil \today\quad\militarytime}
        \let\@evenhead\@oddhead
        \let\label=\draftlabel
        \let\marginnote=\draftmarginnote
        \def\ps@empty{\let\@mkboth\@gobbletwo
        \def\@oddfoot{\hfil \smash{\Large\sl DRAFT} \hfil}
        \let\@evenfoot\@oddhead}
        \def\@eqnnum{(\theequation)\rlap{\kern\marginparsep\tt\@eqnlabel}%
        \global\let\@eqnlabel\@vacuum}  }

\begin{document}



\hfill UTHET-01-0502

\vspace{-0.2cm}

\begin{center}
\Large
{ \bf Thermodynamics of Conformal Field Theories and Cosmology\footnote{Presented at PASCOS 2001,
Chapel Hill, April 2001}}
\normalsize

\vspace{0.6cm}

\baselineskip=14pt
{\bf G. Siopsis}\footnote{gsiopsis@utk.edu}
\\ Department of Physics
and Astronomy,
The University of Tennessee, Knoxville,\\
TN 37996 - 1200, USA.
 \end{center}

\vspace{0.2cm}
\baselineskip=12pt

{\bf Abstract:}
{\em We study the ratio of the entropy to the total energy in conformal field
theories at
finite temperature. For the free field realizations of ${\cal N}=4$
super Yang-Mills theory in $D=4$ and the $(2,0)$ tensor
multiplet in $D=6$, the ratio is bounded from above. The corresponding bounds
are  less stringent  
than the recently proposed Verlinde
bound. For
strongly coupled CFTs with AdS duals, we show that the ratio obeys the
Verlinde bound even in the presence of rotation. For such CFTs, we
point out an
intriguing resemblance in their thermodynamic formulas with 
the corresponding ones of two-dimensional
CFTs. The discussion is based on {\tt hep-th/0101076}~\cite{ego}
}


\normalsize
\baselineskip=14pt\parskip=10pt

\section{Introduction}
The Bekenstein bound \cite{Bekenstein} for the ratio of the entropy
$S$ to the total energy $E$ of a closed physical system that fits in
a sphere in three spatial dimensions reads ~\cite{Veneziano}
\be
\label{Bbound}
\frac{S}{2\pi RE} \leq 1,
\ee
where $R$ denotes the radius of the sphere.
Despite many efforts, the microscopic origin of the bound  
remains elusive. A recent interesting development is Verlinde's
observation \cite{Verlinde} that
CFTs possessing AdS duals satisfy a version of the bound
(\ref{Bbound}). One firstly observes that for general CFTs on
${\mathbb R} \times S^{D-1}$, 
with the radius of $S^{D-1}$ being $R$, the product $ER$ is
independent of the total spatial volume $V$. If one defines
the sub-extensive part of the total energy through
\be
\label{subext}
E_C = DE -(D-1)TS\,,
\ee
then for strongly coupled CFTs with AdS
duals the entropy is given by a generalized Cardy formula
\be
\label{verlentr}
S=\frac{2\pi R}{D-1}\sqrt{E_C(2E-E_C)}\,.
\ee
From (\ref{verlentr}) one obtains a bound similar to (\ref{Bbound}),
namely
\be
\label{BVerl}
\frac{S}{2\pi R E}\leq \frac{1}{D-1}\,.
\ee
In view of the above developments, a natural question arising is
whether there exists a
microscopic derivation of Verlinde's formula (\ref{verlentr}) within
the thermodynamics of CFTs. This question could be checked in the
context of CFTs whose  
microscopic thermodynamics is well understood, such as free CFTs on
${\mathbb R} \times S^{D-1}$~\cite{KL}.
We shall discuss free CFTs in dimensions $D=4,6$ as well as strongly coupled CFTs with AdS duals.

\section{CFTs at weak coupling}


In this section, we discuss the thermodynamics of conformal field theories.
In two dimensions, the entropy and energy of $C$ free bosons read, respectively
\be\label{eq7}
S = {\pi C\over 6}\; \delta^{-1}\,,\quad
ER = {C\over 24} \; \left( \delta^{-2} + 1\right)\,.
\ee
Eq.~(\ref{eq7}) implies the Cardy formula \cite{Cardy3}
\be
S = 2\pi \sqrt{{C\over 6}\; \left( E - {C\over 24} \right)}\,,
\ee
and the Bekenstein bound (\ref{Bbound}) for the ratio
\be
{S\over 2\pi ER} = {2\delta\over 1 +\delta^2} \le 1\,.
\ee
The above results also hold for fermions.

In four dimensions, the ratio $S/E$ in general diverges.
Remarkably, for the ${\cal N}=4$ SYM model, we have
\be
{S\over 2\pi ER} = {2\over 3}\; {2\delta\over 1+\delta^2}\,.\label{SE44}
\ee
There is a critical point at which both $S$
and $ER$ vanish, $\delta_c^2 = 1/3$.
For $\delta\le \delta_c$, we obtain
\be
{S\over 2\pi ER} \le {\sqrt 3\over 3}\,,
\ee
which is weaker than the Verlinde bound (\ref{BVerl}), $S/(2\pi ER)
\le 1/3$. 

It is perhaps worth mentioning that if one imposes periodic boundary
conditions on the gaugino, as suggested by Tseytlin to
account for the 
disagreement on the number of degrees of freedom between the weak and strong
coupling regimes, the above results still hold.

For the $(2,0)$ tensor multiplet in $D=6$, we obtain
\be
{S\over 2\pi ER} = {2\delta\over 1+\delta^2}\; {{3\over 5}-{10\over 3}\delta^2+19\delta^4
\over 1-6\delta^2 + 25\delta^4}\,.
\ee
This is a well-behaved function of $\delta$. We obtain
conclude that
\be
{S\over 2\pi ER} \le 0.824\,,
\ee
which is less stringent than the Verlinde bound (\ref{BVerl}).

\section{CFTs at strong coupling}


In this section we turn our attention to strongly coupled CFTs in
$D$-dimensions, possessing AdS duals. The thermodynamics of such
theories follows quite generally from the thermodynamics of
$(D+1)$-dimensional AdS black holes, through holography. 
We consider the rotating Kerr-AdS (KAdS) black hole in $(D+1)$-dimensions.
According to the AdS/CFT duality conjecture~\cite{Witten1}, the
thermodynamical 
quantities are associated to a strongly coupled $D$-dimensional CFT
residing on the conformal boundary of spacetime,
i.~e.~on a rotating Einstein universe.

Defining $\D = R/r_+$, where $R$ is the AdS throat size and $r_+$ is the horizon radius, and the Bekenstein entropy $S_B = 2\pi ER/(D-1)$,
we obtain after some algebra
\begin{equation}
\frac{S}{S_B} = \frac{2\D}{1+\D^2} \le 1.
\end{equation}
The Bekenstein bound is saturated at the Hawking-Page
transition point.\cite{HP}
We further note that we can write
\begin{equation}
2ER = \frac{D-1}{2\pi}S\frac{R}{r_+}[\D^{-2} + 1]\,, \label{2ER}
\end{equation}
which, for arbitrary $D$, is exactly the behavior of a two-dimensional
CFT (\ref{eq7}) with
characteristic scale $R$, temperature $\tilde T = 1/(2\pi R\D) =
r_+/(2\pi R^2)$,
and central charge proportional to $S_C = SR/r_+$ (Casimir entropy).
Thus, $S_C$ is proportional to the number of degrees of
freedom coupled at the critical point.

Finally, if we define the "Bekenstein-Hawking energy"
$E_{BH}$ as the energy for which the black hole entropy $S$ and
the Bekenstein entropy $S_B$ are equal,
one checks that $E_{BH} \le E$. Furthermore, above the Hawking-Page transition point,
\begin{equation}
E_C \le E_{BH} \le E, \qquad S_C \le S \le S_B,
\end{equation}
where equality holds when the HP phase transition is reached.
As the entropy $S$ is a monotonically increasing function of $E_C$
(or, equivalently, of $r_+$), the maximum entropy is reached when
$E_C = E_{BH}$, i.~e.~at the HP phase transition.
It is quite interesting to observe that at this point the central charge
$c/12 = S_C/(2\pi)$ takes e.~g.~for $D=4$, ${\cal N}=4$ $U(N)$ SYM
theory the value\footnote{Here we used the AdS/CFT dictionary 
$N^2 = \frac{\pi R^3}{2G_5}$.} $c=6N^2$. This is exactly the central charge
of a two-dimensional free CFT containing the $6N^2$ scalars of
$D=4$, ${\cal N}=4$ SYM.

\section{Cosmological implications}

In this section, we discuss the implications of our results for cosmology. The metric in
a radiation
dominated closed Friedman-Robertson-Walker
(FRW) universe,
\begin{equation}
ds^2 = -d\tau^2 + {\cal R}^2(\tau)d\Omega^2_{D-1}, \label{FRW}
\end{equation}
where ${\cal R}(\tau)$ represents the radius of the universe at
a given time $\tau$, is conformally
equivalent to
\begin{equation}
d\tilde s^2 = -dt^2 + R^2d\Omega^2_{D-1}, \label{confmetric}
\end{equation}
where $dt = R\,d\tau/{\cal R}(\tau)$. If the radiation is described
by a CFT, one can equally well use (\ref{confmetric}) instead of (\ref{FRW}).
If, in addition, this CFT admits an AdS dual, it can be described by
a Schwarz\-schild-AdS black hole at some temperature $T$, because
(\ref{confmetric}) is precisely the metric on the conformal boundary
of spacetime.
The observations made by Verlinde \cite{Verlinde} concerning
entropy, energy and temperature bounds in a radiation dominated
universe then fit nicely into this AdS black hole description.
In particular, the universe is weakly (strongly) self-gravitating
if $H{\cal R} \le 1$ ($H{\cal R} \ge 1$), where $H=\dot{{\cal R}}/{\cal R}$
denotes the Hubble constant, and the dot refers to differentiation
with respect to $\tau$. One has $H{\cal R} = 1$
iff the Bekenstein-Hawking entropy $S$ equals the Bekenstein
entropy $S_B$. We saw above that this happens precisely at the
HP transition point $r_+ = R$, so the borderline between the
weakly and strongly self-gravitating regime is the Hawking-Page
phase transition temperature $T_{HP} = (D-1)/2\pi R$. This identification
makes indeed sense, because below $T_{HP}$ (weakly gravitating)
one has AdS space filled with thermal radiation which collapses
above $T_{HP}$ ($r_+ \ge R$, strongly gravitating) to form a black
hole. Furthermore, Verlinde~\cite{Verlinde} found a limiting temperature
for the early universe,
\begin{equation}
T \ge T_H = -\frac{\dot{H}}{2\pi H} \qquad \mbox{for} \quad H{\cal R} \ge 1.
\end{equation}
We conclude that Verlinde's limiting temperature $T_H$
corresponds to the temperature $T_{HP}$ where the HP phase transition
takes place.

\section{Conclusions}

Concerning further developments of our ideas, it might be interesting
to further study our {\it generalized central 
charge} $S_C$, which intriguingly resembles a standard 
two-dimensional central charge. Such an interpretation leads to the
conjecture that there might exist a two-dimensional CFT model whose
dynamics in the presence of irrelevant operators underlies 
the dynamics of the $D$-dimensional CFTs possessing AdS duals. Such a
conjecture might explain the fact that the latter theories
share unexpectedly many of the properties of two-dimensional CFTs.

\section*{Acknowledgments}
Research supported in part by the US Department of Energy under grant
DE--FG05--91ER40627.

\end{document}